\documentclass[a4paper]{jpconf}

\usepackage{amsmath,amssymb}
\usepackage{graphicx,subfig}

\begin{document}
\title{Fractional quantum Hall states in charge-imbalanced bilayer systems}
\author{N. Thiebaut$^1$, N. Regnault$^{2,3}$ and M.O. Goerbig$^1$}
\address{$^1$ Laboratoire de Physique des Solides, CNRS UMR 8502, Univ. Paris-Sud, F-91405 Orsay cedex, France}
\address{$^2$ Laboratoire Pierre Aigrain, ENS and CNRS, 24 rue Lhomond, 75005 Paris, France}
\address{$^3$ Department of Physics, Princeton University, Princeton, NJ 08544}
\ead{thiebaut@lps.u-psud.fr}

\begin{abstract}

We study the fractional quantum Hall effect in a bilayer with charge-distribution imbalance induced, for instance, by a bias gate voltage. 
The bilayer can either be intrinsic or it can be formed spontaneously in wide quantum wells, due to the Coulomb repulsion between electrons. 
We focus on fractional quantum Hall effect in asymmetric bilayer systems at filling factor $\nu=4/11$ and show that an asymmetric Halperin-like 
trial wavefunction gives a valid description of the ground state of the system.

\end{abstract}

\section{Introduction}

In a two-dimensional electron gas subjected to a high magnetic field repulsively interacting
electrons may collectively form a gapped state insensitive to small perturbations. Such states are called fractional quantum Hall states (FQHS) 
and show up for partial (fractional) fillings of the highest partially occupied Landau level, i.e. for rational values of the filling factor $\nu$, 
which is the ratio of the number of electrons $N$ and the number of magnetic flux quanta $N_{\phi}$ threading the system. 
Since the discovery of the fractional quantum Hall effect (FQHE) in 1982 \cite{Tsui} new experimental setups nourished our understanding of 
this multifaceted phenomenon. 

\begin{figure}[h]
\begin{center}
\includegraphics[width=10cm]{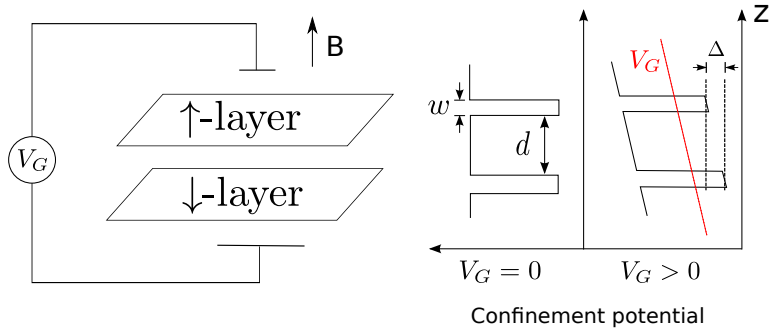}
\caption{Schematic view of the effect of a gate voltage on a bilayer system. \textit{(Left)} Setup. \textit{(Right)} Effect of the gate voltage
on the confinement potential. We consider the limit where the width $w$ of each quantum well is small as compared to the layer separation $d$, $w\ll d$.}
\label{unbalanced_bilayer}
\end{center}
\end{figure}

The bilayer system, which consists of two parallel 2D electron systems in a semiconductor heterostructure exhibits FQHE at even-denominator 
filling factors (such as $\nu=1/2$ \cite{PhysRevLett.68.1383}) 
is one of them. Though it is more common 
to consider symmetric layers, for which charges are equally spread between the two layers ($\nu_{\text{tot}}=2/n$), we are interested in a bilayer 
system with charge imbalance, which is obtained by the application of an external gate voltage (fig. \ref{unbalanced_bilayer}). The strength of 
correlations between electrons that belong to different layers depends on their distance $d$ and the energy difference between the two layers 
$\Delta$ is proportional to the external gate voltage $V_G$. 

Beside its own richness, the FQHE in a bilayer system provides a model for its counterpart in a wide quantum 
well~\cite{PhysRevLett.69.3551,PhysRevLett.101.266804,Papic}, where even-denominator states have also been 
observed~\cite{PhysRevLett.69.3551,PhysRevLett.101.266804,PhysRevLett.103.256802}.
In the latter, density functional theory calculations indicate that the Coulomb repulsion favors an electron concentration at the borders of the 
quantum well, which may thus be viewed as a spontaneously formed bilayer system~\cite{PhysRevLett.108.046804}. Charge-imbalanced bilayers can thus 
depict wide quantum wells in which electrons split in two layers and populate one of them preferentially, either spontaneously or under the influence 
of external gate electrodes. 

Here, we investigate the conditions for the existence of FQHS (at $\nu=4/11$) in an imbalanced bilayer system by means of exact diagonalization. 
We introduce asymmetric Halperin states and discuss their relevance for the description of the ground state of the system.

\section{Model}

We restrict the electron dynamics to the lowest Landau level, which amounts to considering the limit in which adjacent Landau levels are separated by
a large cyclotron gap, as compared to the characteristic Coulomb energy scale, $e^2/4\pi l_B$.
FQHS are often fully polarized due to the Coulomb repulsion between electrons in a flat Landau energy level, and we consider such states, that is we 
neglect the electron spin from now on. The effective Hamiltonian of the bilayer system reads
\begin{equation}\label{Hamiltonian}
\hat{H} =  \frac{1}{2} \sum_{\{\sigma_i\}} \sum_{\{m_i\}} V_{\{m_i\}}^{\{\sigma_i\}} c_{m_1,\sigma_1}^{\dagger} c_{m_2,\sigma_2}^{\dagger} c_{m_4,\sigma_4} c_{m_3,\sigma_3} + \Delta \sum_m (c_{m,\downarrow}^{\dagger} c_{m,{\downarrow}}-c_{m,\uparrow}^{\dagger} c_{m,{\uparrow}}) \quad ,
\end{equation}
where $\sigma=\uparrow,\downarrow$ is the layer index and $m_i$ the angular momentum which labels the single-electron states in the symmetric 
gauge\footnote{Here we assume that the confinement potential is sufficiently narrow (compared to the layer separation, $w\ll d$) to avoid 
electrons from populating the first excited state of the confinement potential. This permits to neglect the wavefunction extent 
in the confinement direction, which assumption simplifies the single particle Hilbert space as the layer index does not couple to the orbital motion.} 
\cite{Girvin}; $\Delta$ is the energy difference between the two layers, controlled by the gate voltage $V_G$ in fig.~\ref{unbalanced_bilayer}. 
The first term on the right hand side of eq.~\ref{Hamiltonian} is the usual interaction term, while the second one is a Zeeman-like term 
proportional to the density imbalance. In real space the interaction potential depends on the layer indices
\begin{equation}\label{interaction}
 V^{\{\sigma_i\}}(r)=  \left\{ \begin{array}{cl} \frac{e^2}{4\pi r} & \mbox{ if electrons belong to the same layer (i.e. if }  
\sigma_1=\sigma_2= \sigma_3 =\sigma_4\text{)}\\ & \\
 \frac{e^2}{4\pi \sqrt{r^{\,2}+d^{\,2}}} & \mbox{ if electrons belong to different layers (i.e. if } \sigma_1=\sigma_3 \neq \sigma_2 =\sigma_4 \text{)}\\
 & \\
 0 & \mbox{elsewise}
\end{array} \right . 
\end{equation}

\section{Halperin's wavefunctions}

A large variety of FQHS in bilayer systems may be described with the help of a generalization of Laughlin's wavefunctions~\cite{Laughlin} 
to the case of electrons with internal degrees of freedom, initially introduced by Halperin~\cite{Halperin}. Those wavefunctions have three 
integer parameters that characterise the interaction strength between electrons in the $\uparrow$ and $\downarrow$ layers 
($m_{\uparrow}$ and $m_{\downarrow}$), and electrons in different layers ($n$). 
In terms of the complex positions $z_i^{\sigma}=x_i^{\sigma}-i y_i^{\sigma}$, for a system with $N_{\sigma}$ electrons in the $\sigma$-layer, 
Halperin's wavefunctions read
\begin{equation}\label{Halperin}
\varPsi_{(m_{\uparrow}, m_{\downarrow}, n)}\big( \{z_i^{\uparrow}\},\{z_i^{\downarrow}\} \big) = 
  \prod_{i<j\leq N_{\uparrow}} (z_i^{\uparrow} - z_j^{\uparrow})^{m_{\uparrow}} \prod_{i<j\leq N_{\downarrow}} (z_i^{\downarrow} - z_j^{\downarrow})^{m_{\downarrow}} \prod_{i\leq N_{\uparrow} , \, j\leq N_{\downarrow}}   (z_i^{\uparrow} - z_j^{\downarrow})^n \quad ,
\end{equation}
where the normalization and usual Gaussian factors have been absorbed in a redefinition of the Hilbert space measure 
$d\mu (\{z_i\}) = \prod_i dx_i dy_i \exp(-|z_i|^2/2 l_B^2)/(2 \pi l_B^2) $, following Girvin and Jach~\cite{PhysRevB.29.5617}. Alternatively, 
we denote the state represented by the Halperin wavefunction~(\ref{Halperin}) the $(m_{\uparrow}, m_{\downarrow}, n)$ state.
Notice that not all Halperin state~(\ref{Halperin}) are capable of describing physically possible FQHS -- it has been shown in Ref.~\cite{PhysRevB.77.165310}
within a generalization of Laughlin's plasma analogy~\cite{Laughlin} that the exponents must satisfy the condition 
\begin{equation}\label{condition}
 m_{\uparrow}m_{\downarrow}\geq n^2
\end{equation}
in order to avoid phase separation of the electron species.

To compute the filling factor associated with a particular Halperin state, one uses the fact that the maximum value of individual orbital momenta is 
given by the number of flux quanta $N_{\phi}$ threading the system. The polynomial expansion of eq.~(\ref{Halperin}) yields a maximal exponent 
$( N_{\sigma} - 1 ) \; m_{\sigma} + N_{-\sigma}\; n$ for the position of the $i$-th electron in the $\sigma$-layer that must equal $N_{\phi}$.
This can be written in matrix form as

\[
\begin{pmatrix} N_{\phi}+m_{\uparrow}\\ N_{\phi}+m_{\downarrow} \end{pmatrix}
=
\begin{pmatrix}
m_{\uparrow} &  n \\ 
n &  m_-{\downarrow}
\end{pmatrix}
\begin{pmatrix} N_{\uparrow} \\ N_{\downarrow} \end{pmatrix} \quad ,
\]
and after inverting the matrix, one deduces the relationship between the total number of electrons $N$ and $N_{\phi}$,
\begin{equation}\label{shift}
N=N_{\uparrow}+N_{\downarrow}=\frac{1}{m_{\uparrow}m_{\downarrow}- n^2} \Big[ (m_{\uparrow}+m_{\downarrow}-2n) N_{\phi} + 2 m_{\uparrow} m_{\downarrow} - n(m_{\uparrow}+m_{\downarrow})\Big] \quad .
\end{equation}
The filling factor $\nu$ is defined in the thermodynamic limit, for which we obtain
\begin{equation}
 \nu = \lim_{N,N_{\phi}\rightarrow \infty} \frac{N}{N_{\phi}} =\frac{m_{\uparrow}+m_{\downarrow}-2n}{m_{\uparrow}m_{\downarrow}- n^2} \quad .
\end{equation}
We define the polarization as the population difference between the layers
\begin{equation}\label{pol}
 P_z=\frac{S_z}{N/2}
\end{equation}
in terms of the $z$-component of the total layer pseudospin
\begin{equation}\label{pseudospin}
S_z = \frac{N_{\uparrow}-N_{\downarrow}}{2} = \frac{(m_{\downarrow}-m_{\uparrow})(N_{\phi}-n)}{2(m_{\uparrow}m_{\downarrow}-n^2)} \quad ,
\end{equation}
which is zero for symmetric states (i.e. if $m_{\uparrow}= m_{\downarrow}$). 

Now we focus on asymmetric Halperin states. More precisely we are interested in the $(5,3,2)$ state, which satisfies the condition (\ref{condition})
since it is the simplest 
(i.e. lowest correlation factors) asymmetric Halperin state for which the filling factor $\nu_{532}=4/11$ does not belong to the usual 
composite-fermions sequence $\nu_{\text{CF}}=p/(2sp\pm 1)$ \cite{Jain_CF}. It may thus describe states the existence of which relies on layer asymmetry. 
Nevertheless it should be mentioned that $\nu=4/11$ is also the filling factor of a second-generation composite fermion state in 
monolayer systems~\cite{Pan_FQHEOfCF,PhysRevB.69.155324,PhysRevB.69.155322,PhysRevLett.92.196806} generated by interacting composite fermions
that has been proposed as a possible explanation for an experimentally observed FQHS at this filling \cite{Pan_FQHEOfCF}.

\section{Exact diagonalization}

In order to check the physical relevance of the $(5,3,2)$ state we calculate, with the help of exact diagonalization~\cite{DiagHam}, 
two quantities for the Coulomb interaction (\ref{interaction}) in the lowest Landau level: 
the total angular momentum of the ground state and its polarization as a function of the distance $d$ and 
of the gate-induced energy difference $\Delta$ between the two layers. The calculations are perfomed on the sphere geometry~\cite{Haldane}. For the $(5,3,2)$ state, the relation between $N_\Phi$ and $N$ is given by $N_\Phi=\frac{11}{4} N - \frac{7}{2}$. Finite-size systems induce a constraint on the number of electrons 
[see eq.~(\ref{shift})]. Only sizes of the form $N=2+4n$, $n\in \mathbb{N}$ can satisfy this constraint. 

Here we will focus on $N=6$ and $N=10$ which are the only accessible sizes. 
FQHS correspond to values of $d$ and $\Delta$ for which the gap is finite and the total angular momentum 
equals zero. Finally we compute the overlap between the $(5,3,2)$ state and the exact ground state for values of $d$ and $\Delta$ which correspond 
to a FQHS. 

\begin{figure}[h]
\centering
\subfloat[N=6: $P_z^{(532)}=-1/3$ ]{\includegraphics[width=7cm]{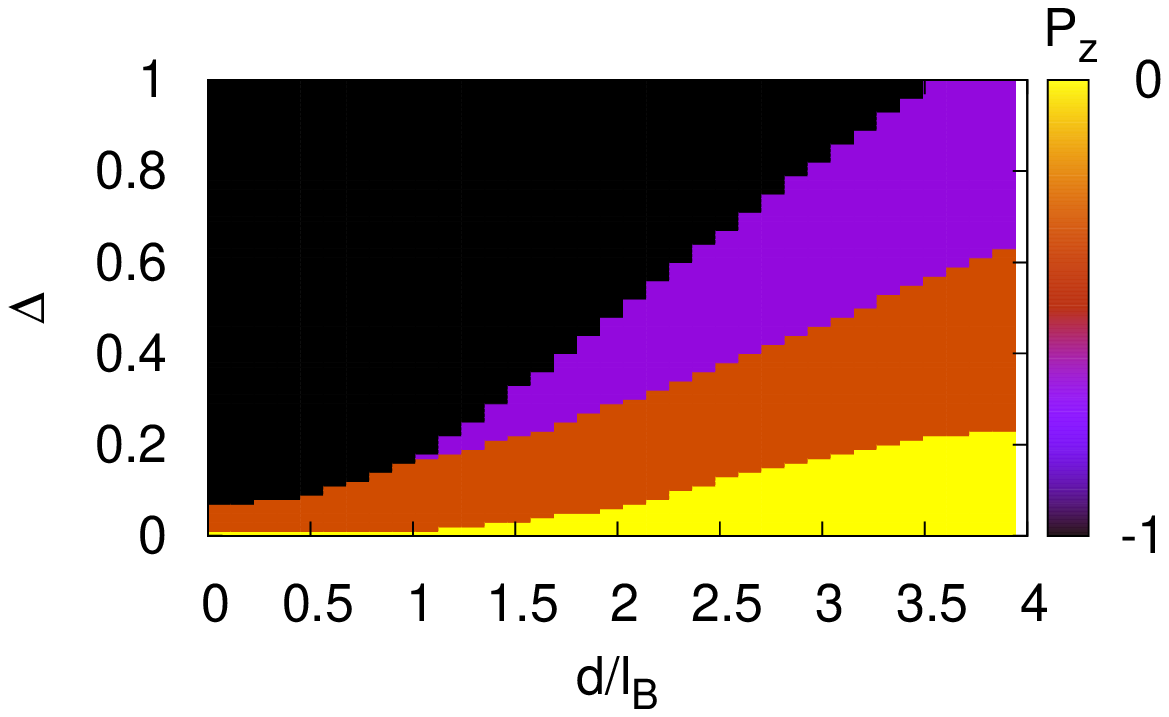}}
\subfloat[N=10: $P_z^{(532)}=-2/5$]{\includegraphics[width=7cm]{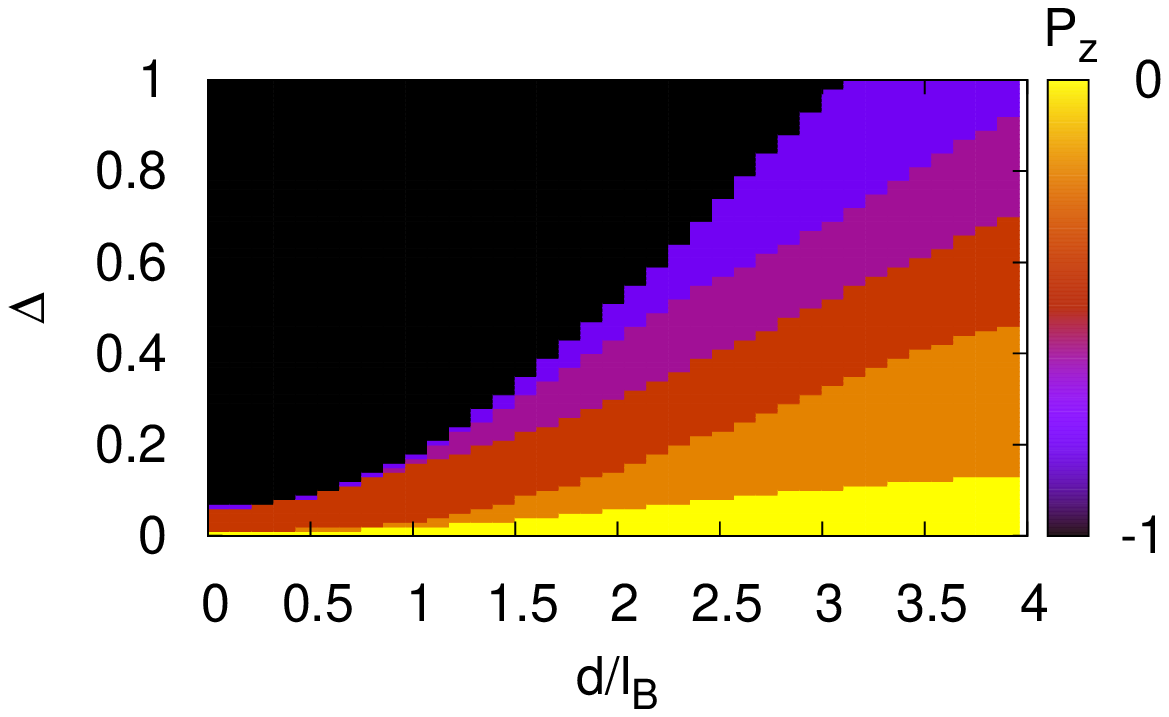}}
\caption{Polarization of the bilayer system as a function of the distance $d/l_B$ and energy shift $\Delta$ (energies are given in units 
of $e^2/4\pi l_B$).} 
\label{polarization}
\end{figure}

The polarization of the ground state is shown in fig.~\ref{polarization}. For sufficiently high values of $\Delta$ the system is fully polarized, 
i.e all electrons reside in a single layer and one thus obtains the monolayer system with $\nu=4/11$, studied in Refs.~\cite{PhysRevB.69.155324,PhysRevB.69.155322,PhysRevLett.92.196806}. Conversely for large distances $d$ the polarization tends to be zero. 
In this situation the potential $V(r)=(r^{\,2}+d^{\,2})^{-1/2}$ between two electrons in different layers is small and the system behaves as 
two uncorrelated layers with individual filling factors $\nu=2/11$. It is likely that for such a low filling factor the system tends to form a 
Wigner crystal~\cite{PhysRevB.30.473,PhysRevLett.65.633,PhysRevB.44.8107}, and FQHS may therefore be ruled out. 

The polarization of the $(5,3,2)$ state can be computed directly from eq.~(\ref{pol}) and eq.~(\ref{pseudospin})
\begin{equation}\label{eq:pol}
P_z^{(532)}=-\frac{1}{2}+\frac{1}{N}
\end{equation}
and may be compared with that obtained from exact diagonalization for $N=6$ and 10 particles, as a function of the 
layer separation $d/l_B$ and $\Delta/(e^2/4\pi l_B)$ (see fig.~\ref{polarization}). Indeed, one finds that in intermediate regimes the polarization of the ground state is the 
one of the $(5,3,2)$ state (orange regions).

\begin{figure}[h]
\centering
\includegraphics[width=7cm]{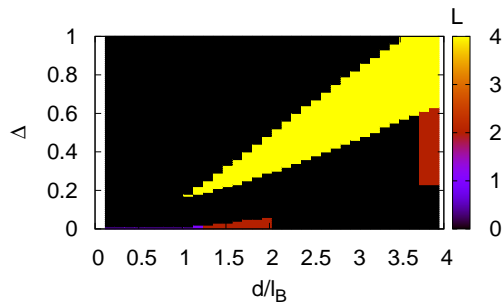}
\caption{Total orbital momentum of the bilayer system for $N=6$ electrons as a function of the distance $d/l_B$ and energy shift $\Delta$,
in units of $e^2/4\pi l_B$.}
\label{momentum}
\end{figure}

We now turn to the characterization of the exact ground state. Figure \ref{momentum} shows the total angular momentum $L$ of the ground state. This latest is rotationally invariant when $L=0$, as required for an incompressible state. It is zero in the region where the ground-state polarization matches that of the $(5,3,2)$ state, 
except for large layer separations. As already mentioned in the discussion above, 
for large distances inter-layer correlations are too weak to allow the system to be in an incompressible state, and an inhomogeneous density state 
is favored, such as one would for example expect for a Wigner crystal ($d \gtrsim 3.8 l_B$).

\begin{figure}[h!]
\centering
\includegraphics[width=6.5cm]{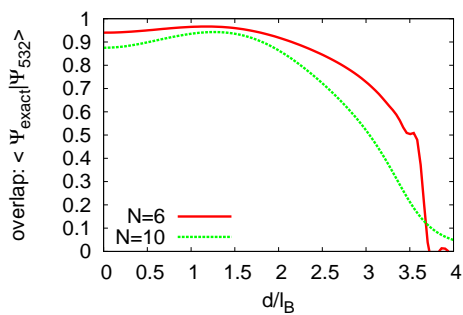}
\caption{Overlap of the $(5,3,2)$ state with the exact ground state as a function of $d/l_B$, in the corresponding polarization sector (\ref{eq:pol}).}
\label{overlap}
\end{figure}

The results presented in figs.~\ref{polarization} and~\ref{momentum} indicate that in a certain range of parameters $d/l_B$ 
and $\Delta/(e^2/4\pi l_B)$, exact diagonalization provides a ground state that matches some physical properties of the $(5,3,2)$ state. 
To corroborate the relevance 
of this state, we have calculated the overlap between the exact ground state obtained for the interaction~(\ref{interaction}) and the $(5,3,2)$ state.
In order to obtain the $(5,3,2)$ state numerically, we have used exact diagonalization for a model interaction given in terms of the 
appropriate pseudopotentials~\cite{Haldane}. 
The results are presented in fig.~\ref{overlap} as a function of $d/l_B$. The overlap turns out to be high ($>85\%$) for $d\lesssim 2l_B$, such that in regions of matching polarizations and for sufficiently short interlayer distances the $(5,3,2)$ state gives a good description of the ground state.

\newpage
\section{Conclusion}

We have performed an exact-diagonalization study of the bilayer system for various distances between the layers and energy shifts, i.e. the energy difference between electrons in the two layers. The ground state is incompressible 
over a certain range of parameters, which corresponds to an interlayer distance of roughly three times the magnetic length. Asymmetry 
between layers does not necessarily destroy the FQHS, and the FQHE at exotic fractions may thus be observed in an asymmetric bilayer system. 
Asymmetric Halperin's  wavefunctions are good candidates for the accurate description of those states. Also, using the bilayer modeling of a wide 
quantum well, a similar behavior could be expected in asymmetric wide quantum wells, the asymmetry being induced, for instance, by a back gate 
voltage that enables one to tune the electronic density of the system. Experimentalists reported a FQHE at $\nu=4/11$ in a wide 
quantum well ($w\sim 3\; l_B$) in 2003~\cite{Pan_FQHEOfCF}. Since this filling factor corresponds to the one of
the $(5,3,2)$ state, one may wonder about its relevance for the explanation of this experimental observation. Within 
this picture this FQHS would be induced by the shape of the confinement potential, in contrast to the picture of second-generation 
composite fermions that have been proposed as an alternative explanation for 
the 4/11 state~\cite{Pan_FQHEOfCF,PhysRevB.69.155324,PhysRevB.69.155322,PhysRevLett.92.196806}.

\bigskip
\bibliographystyle{iopart-num}
\bibliography{hmf}
\end{document}